# Predictive Synthesis of MOFs and COFs

*Bin Feng,[1,2] Reid A. Coyle,[1] Shyam Chand Pal,[1,2] Peter Walther,[1] Saeun Park,[1] Zhiling Zheng[1,2,*]*

[1]Department of Chemistry, Washington University, St. Louis, Missouri 63130, United States

[2]Institute of Materials Science and Engineering, Washington University, St. Louis, Missouri 63130, United States

KEYWORDS

Chemical synthesis, metal organic frameworks, covalent organic frameworks, machine learning, artificial intelligence

ABSTRACT

Predictive synthesis concerns the prospective modeling of how a targeted structure can be realized. The synthesis of metal-organic frameworks (MOFs) and covalent organic frameworks (COFs) intrinsically involves a multidimensional space of interdependent variables, posing a challenge for bench chemists to parameterize and predict. In this Perspective, we discuss the evolution from chemical-intuition-driven experimentation toward the integration of artificial intelligence (AI) and machine learning (ML) in the laboratory. By virtue of prior knowledge, uncertainty, and experimental failure in determining what should be attempted next, we envision predictive synthesis progressing from static recipe recommendation toward real-time control and transferable principles of framework formation. In this context, the significance lies in transforming trial-and-error practice into a means of measurable chemical reasoning, thereby accelerating crystalline materials discovery and advancing synthetic chemistry.

## Introduction

The discovery of crystalline reticular materials is governed by the linking of molecular building units through strong bonds to form periodic structures,[1] yet the conditions under which such structures emerge comprise many experimental variables that are individually simple but collectively nontrivial.[2–5] Conceptually, both metal-organic frameworks (MOFs) and covalent organic frameworks (COFs) can be designed from nodes and linkers,[6–11] albeit the realization of the intended crystalline phase depends on concentration, modulator identity and amount, solvent environment, time, temperature, and nucleation kinetics.[12–18] In light of this complexity, the most consequential decisions in MOF and COF synthesis are often made not at the selection of framework based on properties alone,[19,20] but at the level of synthesizable materials verification and reaction condition discovery.[21–26] In fact, in contrast to the increasing attention given to *in silico* design of MOFs and COFs,[4,27–29] computer-assisted synthesis of such reticular materials remains at an early stage,[30,31] presenting an opportunity to convert an expanding space of conceivable frameworks into materials that can be realized, studied, and ultimately deployed for societal benefit.[30,32–34]

At the core of this gap lies the so-called "crystallization challenge": *how do reaction conditions govern the formation of a crystalline framework?* Hitherto, this condition-search problem has been approached through chemical intuition,[35–39] literature analogue,[40–45] systematic high-throughput screening,[46–48] and design-of-experiments campaigns.[49–52] However, for most experimentalists, quantitatively evaluating and prospectively identifying promising conditions within a large synthesis space remains challenging. The emergent progress in applying algorithms and generative AI across scientific research offers new solutions.[53–55] In this Perspective, we argue that reticular compounds including MOFs and COFs are uniquely well-suited to benefit from AI-guided predictive synthesis due to two intrinsic properties: (i) *modularity*, which allows systematic variation and increases the abundance of chemically related synthesis data; and (ii) *structural clarity*, which enables clear synthesis-structure-property relationship at single-crystal level. These features make reticular synthesis unusually data-tractable and provide a basis for learning how experimental decisions give rise to crystallization (**Figure 1**).

In this regard, the Perspective will outline the ongoing and future actions to capture chemical intuition in MOF and COF synthesis with decision-making algorithms such as Bayesian optimization[56–58] and genetic algorithms.[59–61] We further consider hybrid AI systems that couple large language models (LLMs) trained on the scientific literature with data-driven approaches, such that experimentally defined objectives—framework crystallinity or stability, for example—can guide the prioritization of synthesis conditions. Incorporating prior knowledge, uncertainty quantification, and historical failures is therefore essential for overcoming optimism bias while improving model interpretability.[62,63] At a fundamental level, AI-guided experimentation has the potential to transform the practice of navigating chemical space of MOFs and COFs into a means of rendering synthesis predictive and measurable, thereby quantifying the influence of reaction conditions and deriving new chemical insights.

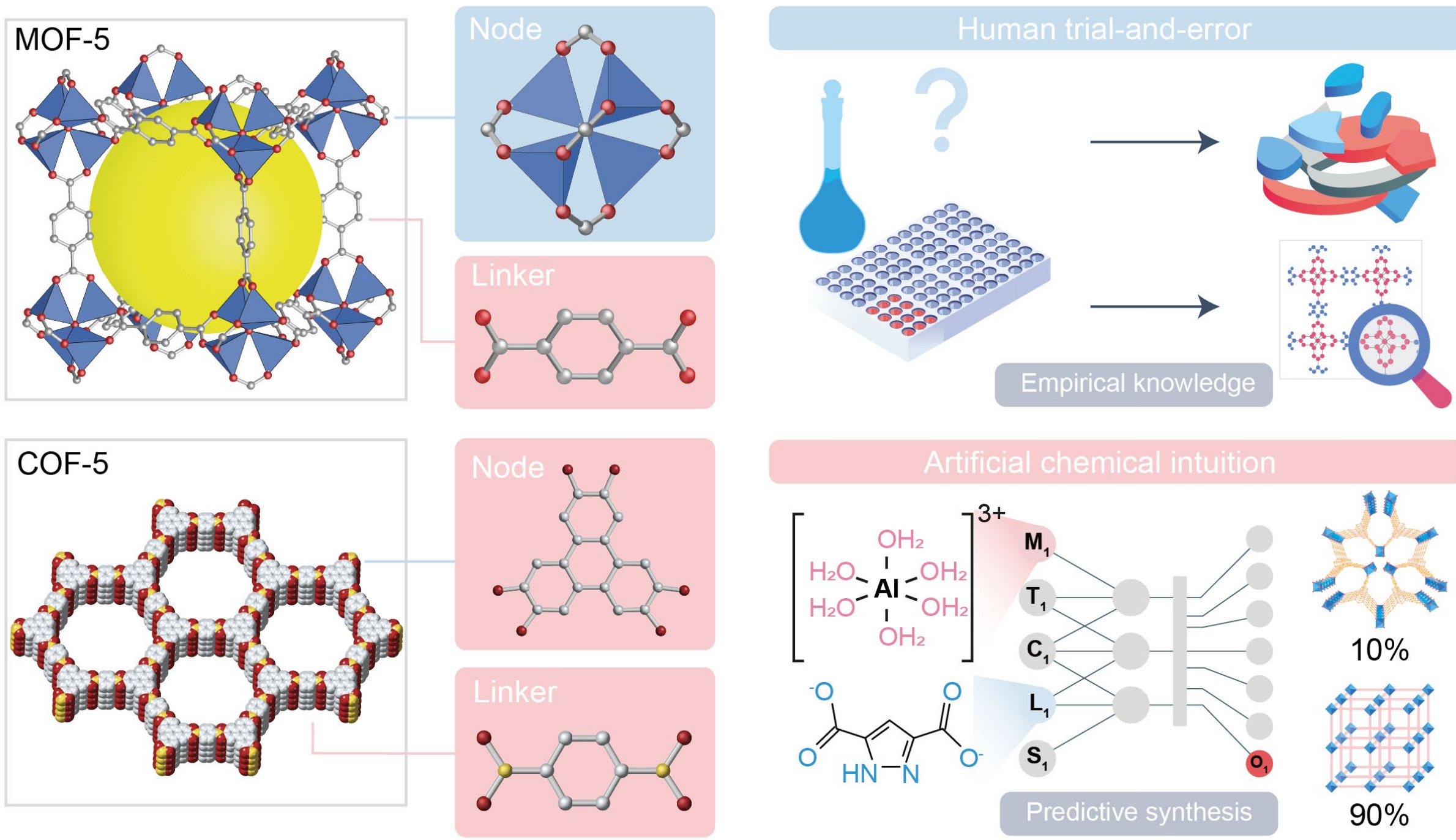

**Figure 1.** Comparison between traditional synthesis guided by human intuition and AI-directed synthesis of MOFs and COFs. Left: modular construction of MOFs and COFs, exemplified by MOF-5 and COF-5. Color code: Zn, blue polyhedra; C, gray; O, red; B, yellow. The yellow sphere represents pore space. Hydrogen atoms are omitted for clarity. Right: manual trial-and-error synthesis (top) and AI-directed synthesis in which chemical intuition is encoded algorithmically (bottom).

**Reticular synthesis as a learnable system**

Over the past three decades, the synthesis of MOFs and COFs has progressed from isolated structures to rationally designed families (**Figure 2a**), in which building block metrics[64–71] and functionality[45,72–76] can be systematically varied while retaining the same underlying topology[9,77–79] through the isoreticular principle—a term derived from the Greek *iso-* ("same") and *reticular* ("netlike").[1,80,81] At the bench, this multiplicity is exemplified by MOF-5 [$Zn_4O(BDC)_3$; $BDC^{2-}$ = benzene-1,4-dicarboxylate][82,83] and its expanded families[64,84] and functionalized derivatives,[74,85] all of which retain the $Zn_4O(–COO)_6$ secondary building unit (SBU) in a primitive cubic arrangement. This hierarchical mode of design and synthesis has since been extended to well-studied systems such as the UiO-66 [$Zr_6O_4(OH)_4(BDC)_6$] series,[41,86–88] the NU-1000 [$Zr_6O_4(OH)_4(TBAPy)_2$; $TBAPy^{4-}$ = 1,3,6,8-tetrakis(p-benzoate)pyrene] series[89–91] and the MOF-74 [$M_2(DOT)$; $M^{2+}$ = Zn, Mg, Mn, Co, or Ni; $DOT^{4-}$ = 2,5-dioxidoterephthalate] series (**Figure 2b**).[92–94]

In this regard, the learnability of reticular synthesis arises from the mutual benefit of two features: (i) modularity, which generates related families of experiments through systematic changes in nodes and linkers, and (ii) structural resolvability, which can provide chemically explicit endpoint labels when phase identity, crystallinity, guest–host interactions, and defects are established rigorously.[3,9,30–32,95,96] Consider the early examples COF-5 [$C_9H_4BO_2$; $(HHTP)_2(BDBA)_3$; boronate ester; HHTP = 2,3,6,7,10,11-hexahydroxytriphenylene; BDBA = 1,4-benzenediboronic acid] and COF-10 [$C_{12}H_6BO_2$; $(HHTP)_2(BPDA)_3$, boronate ester; BPDA = 4,4′-biphenyldiboronic acid].[69,97] Extending BDBA to BPDA preserves the hexagonal sheet (i.e., the same two-dimensional connectivity), while increasing the pore diameter by 5 Å and the pore volume by 0.37 $cm^3\ g^{-1}$, as measured by powder X-ray diffraction (PXRD) and nitrogen sorption.[98] Likewise, TpPa-1 [$(Tp)_2(Pa\text{-}1)_3$, β-ketoenamine; Tp = 1,3,5-triformylphloroglucinol, Pa-1 = *p*-phenylenediamine] established a β-ketoenamine platform in which the diamine controls the chemical environment presented within the pore;[99] replacement by a pyridyl diamine gives COF-323 [$(Tp)_2(DAPy)_3$, β-ketoenamine; DAPy = 2,5-diaminopyridine],[100] with a correspondingly altered hydrophilic environment and water-sorption behavior.[23] Hence, the modular ability to construct precise framework variations, coupled with crystallographically resolved outcomes, gives rise to an abundant design space and expanding structural datasets,[19,101–104] makes MOFs and COFs an unusually disciplined playground for both humans and AI/ML to understand how experimental decisions give rise to framework outcomes.

Structure by design, however, does not yet provide formation by design. There remains a disparity between what can be conceived and what synthetic practice can reliably realize (**Figure 2c**).[4,33,63,105] Hitherto, synthesis records have usually compressed each experiment into a static row containing reagents, nominal conditions, and a final characterization result.[2,30] Even when building-unit geometry is compatible with a target topology, its realization depends on precursor speciation before assembly, error correction during bond formation, as well as growth and phase transformation after nucleation.[106–108] Since the crystallization process requires reversible bond formation to allow self-correction of defects and avoid kinetically trapped products, for MOFs, essentially hydrolysis and deprotonation regulate metal–linker bond formation, whereas for COFs, bond exchange and modulation activity govern covalent error correction.[95,109] A conceivable framework material can therefore remain synthetically inaccessible despite being topologically reasonable, and a reported recipe can prove fragile under subtle changes in synthesis steps, washing procedures, or action profile.[40,110–112] As such, a successful recipe identifies one point in synthesis space; the challenge of predictive synthesis lies in establishing the domain over which the targeted framework can be formed reproducibly.

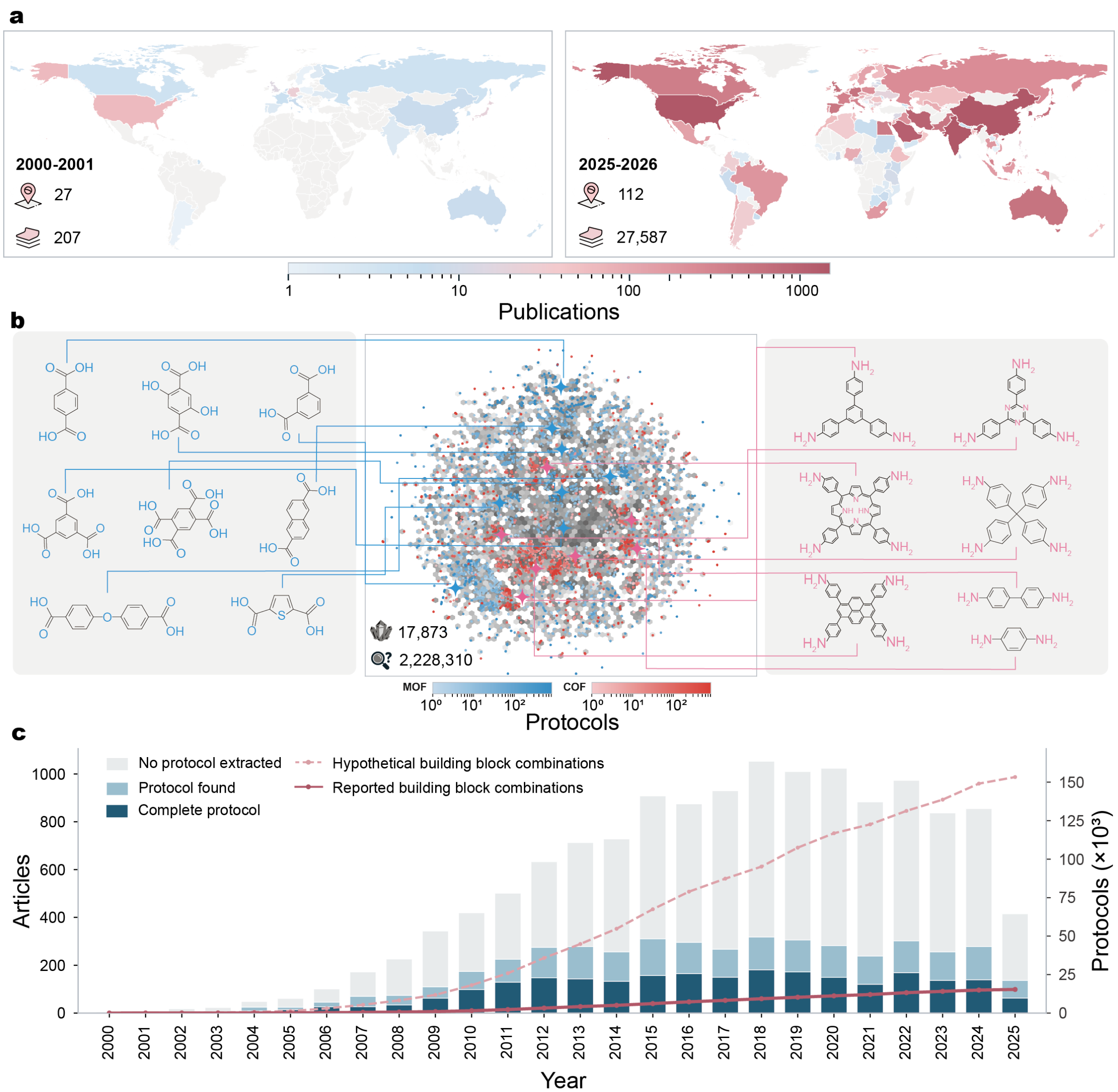

**Figure 2.** Overview of the development of MOF and COF research. (a) Number of publications and geographic distribution of contributing institutions for MOFs and COFs in 2000 and 2026. (b) Uniform manifold approximation and projection (UMAP) comparing reported synthesis protocols for MOFs (blue) and COFs (red) with the potential synthesis space. (c) Fraction of MOF publications reporting complete synthesis protocols among all field-related publications. A protocol is defined as complete when it specifies linker concentration, metal concentration, metal-to-linker ratio, temperature, reaction time, primary solvent, and modulator. Lines represent the cumulative numbers of potential and reported building-block combinations.

In light of this path dependence, we define an experimental synthesis trajectory as the time-ordered sequence of chemical states, actions, observations, and outcomes through which molecular building units become a framework.[30,108,113–115] Such a trajectory of MOFs and COFs comprises (i) an initial state, defined by the reagents, their proportions, and the reaction environment; (ii) experimental actions, including addition, mixing, heating, and activation; (iii) evolving states, where precursors become clusters or oligomers before nucleation and framework growth; (iv) observations, by which these changes are rendered measurable; and (v) an outcome, established with respect to phase identity, porosity, stability, and reproducibility. A representation to show accumulated trajectory at time $t$ can be written as

$$T_t = \{x_0, (a_1, o_1), (a_2, o_2), \dots, (a_t, o_t)\} \quad \textbf{(Eq. 1)}$$

where $x_0$ denotes the initial chemical state, $a_t$ is an experimental action, and $o_t$ is the resulting observation. The next action is selected according to

$$a_{t+1} = \pi(T_t) \quad \textbf{(Eq. 2)}$$

where $\pi$ represents the rule or policy by which the accumulated experimental history informs what is done next, with the goal of separating the recipe, which specifies intended operations, from the trajectory, which records what was actually carried out and observed. For example, two otherwise identical experiments sharing the same nominal composition and temperature may follow different trajectories when the order of addition changes precursor speciation, which in turn alters nucleation.[116–118] In essence, the synthesis trajectory records how its formation became possible and, in this manner, the quality of the decision in predictive synthesis therefore depends on the precision of the observations from which it is made.

Once synthesis is represented as a trajectory, three levels of predictive tasks become distinguishable. Frist, static recipe prediction maps building units and proposed conditions to an endpoint such as phase identity, crystallinity, porosity, or yield, with the experiment compressed into one table row.[14,17,62,119–121] Second, prospective experiment selection uses prior outcomes and quantified uncertainty to choose the next condition set, with objectives including target identification, phase-boundary mapping, and location of a reproducible condition domain.[23,53,122,123] Third, adaptive trajectory control operates during the current MOF and COF synthesis, where time-resolved observations alter subsequent actions based on the feedback control of crystallization (**Figure 3e**).[124–126] It is instructive to distinguish here that synthesis prediction is often referred to as the estimation of the outcome of a specified experiment, whereas predictive synthesis is the prospective selection and control of an experimental synthesis trajectory that realizes a targeted structure with quantified confidence and an account of the variables governing its formation.[30,62,63] Still, predicting the conditions under which a targeted framework will form remains a long-standing challenge in MOF and COF chemistry (i.e., the "crystallization challenge"), since establishing synthesis–structure–property relationships ultimately requires understanding how experimental decisions give rise to a particular structural outcome.

## The Experimental Basis of Predictive Synthesis

The abundance of reported framework structures should not be mistaken for an abundance of synthesis evidence. It is worthwhile to recognize that, while structural repositories count more than $10^6$ deposited reticular structures,[32,127,128] predictive synthesis of MOFs and COFs conceptually requires identifiable experimental attempts that connect an intended framework target with the conditions applied and the outcome experimentally obtained. A useful synthesis-evidence unit therefore joins (i) the building units, linkage, and assigned framework phase; (ii) a complete protocol linked to each identifiable attempt; and (iii) provenance sufficient to distinguish an originating report from an independent replication. There is consequently a disparity between what is structurally known and what is supported by reproducible formation evidence. The magnitude of this disparity, for instance, is reflected in a recent survey of 130 MOFs suggesting that their initial reports most had no independently reported repeat synthesis even after more than a decade following.[129] Thus, one should be aware that growth in the number of structures can coexist with sparse evidence concerning the conditions under which those structures form reproducibly.[130]

In order to assemble connected evidence, the comprehensive knowledge collection of chemical synthesis information from MOF and COF literature reports can be considered as a logical step toward rendering reported protocols and structures both searchable and comparable. Until the mid-2010s, much of this information remained embedded in prose and supporting information, making comparison across thousands of syntheses difficult, while experimental data more broadly remained distributed across formats that were poorly accessible to digital analysis.[131] Early text-mining approaches therefore established that reagents and reaction conditions could be recovered systematically from publications,[18,120] while structured resources subsequently organized such information into searchable synthesis records.[12,17] Since 2023, more recent LLM-based approaches have expanded this capability by extracting synthesis information across heterogeneous writing formats and organizing more than $10^4$ synthesis protocols into machine-readable records and linking them to structure or property data.[13,14,16,132,133] This progression points toward multimodal records in which prose, tables, reaction schemes, and characterization files are associated with the same synthesis event.[16,134,135] Yet knowledge extraction is an act of recovering what was documented, and the challenge now is to preserve the distinction between missing information and observed negative evidence.

Indeed, failure itself must also be defined with respect to the intended framework outcome. For predictive synthesis to advance as a scientific endeavor, failed MOF and COF synthesis attempts are as consequential as successful ones;[23,121,136,137] more fundamentally, not all failures are chemically or statistically equivalent. For instance, a reaction yielding an amorphous solid and one yielding a competing crystalline phase may both miss the intended target, albeit carrying fundamentally different chemical information, as exemplified by studies of Group 4 metal MOFs (e.g., Ti and Zr) and imine-linked COFs.[138–141] Likewise, low or trace crystallinity may indicate incomplete ordering,[96,106] whereas formation of the target followed by collapse during activation implicates a later part of the experimental synthesis trajectory.[110,142] Consequently, framework outcomes can be organized as (i) an unresolved product, including no isolated solid or amorphous material; (ii) a resolved off-target product, including a competing phase or phase mixture; and (iii) an intended-phase product, encompassing (a) crystallinity ranging from low to high and (b) retention or loss of permanent porosity and structural stability upon activation. Such outcome-relative labels preserve chemical meaning and allow partial success to identify boundaries that a binary PXRD-derived crystalline or non-crystalline label table would otherwise erase.

In practice, however, assembling such synthesis evidence remains nontrivial, as the observations most valuable for learning are often those least consistently recorded at the bench. Electronic laboratory records, digitized archival laboratory notebooks, instrument logs, and wearable recording devices provide complementary means of capturing time-stamped experimental actions and observations associated with their source,[30,143–146] and their broader adoption can be advanced through thoughtful integration into routine synthetic practice—comprehensive recording should by no means impose a substantial additional burden on experimentalists, particularly through demands on bench time that make consistent documentation difficult to sustain.[147–151] It is conceivable that an "agent-in-the-laboratory" could assist in this regard; in other words, an AI system operating alongside the chemist could collect heterogeneous experimental records and organize them into uniform, machine-readable representations as the experiment is carried out.[32,152–154] In this manner, success, failure, provenance, and uncertainty become connected across individual synthesis attempts, giving rise to a map in which reproducible domains, competing phases, and uncertain regions can be distinguished. Having such evidence in hand, the next logical step is to determine how these accumulated observations can be translated into quantitative experimental decisions, such that each new synthesis is selected for the information it contributes toward controlling framework formation.

## Capturing Chemical Intuition with Algorithms

In laboratory synthesis of MOFs and COFs, chemical intuition is inherently manifested in the sequence of choices by which a chemist (i) varies reaction parameters, (ii) interprets an outcome, and (iii) decides what should be attempted next. It is useful to recognize that such intuition is accumulated unevenly across synthesis space.[155–157] In this sense, "success" provides points of reference, whereas "failure" defines boundaries beyond which a chosen chemical path no longer gives the intended framework.[30] Thus far this intuition has been indispensable because precursor chemistry, bond formation, solvent environment, and nucleation are coupled in ways that make their individual effects difficult to disentangle.[107,111,156] Yet the very quality that makes chemical intuition valuable—its accumulation through experience and observation—also makes it difficult to quantify, compare, and transfer. To this end, algorithms provide a means of rendering these choices measurable by relating experimental variables to observed outcomes and the uncertainty associated with them (**Table 1**).

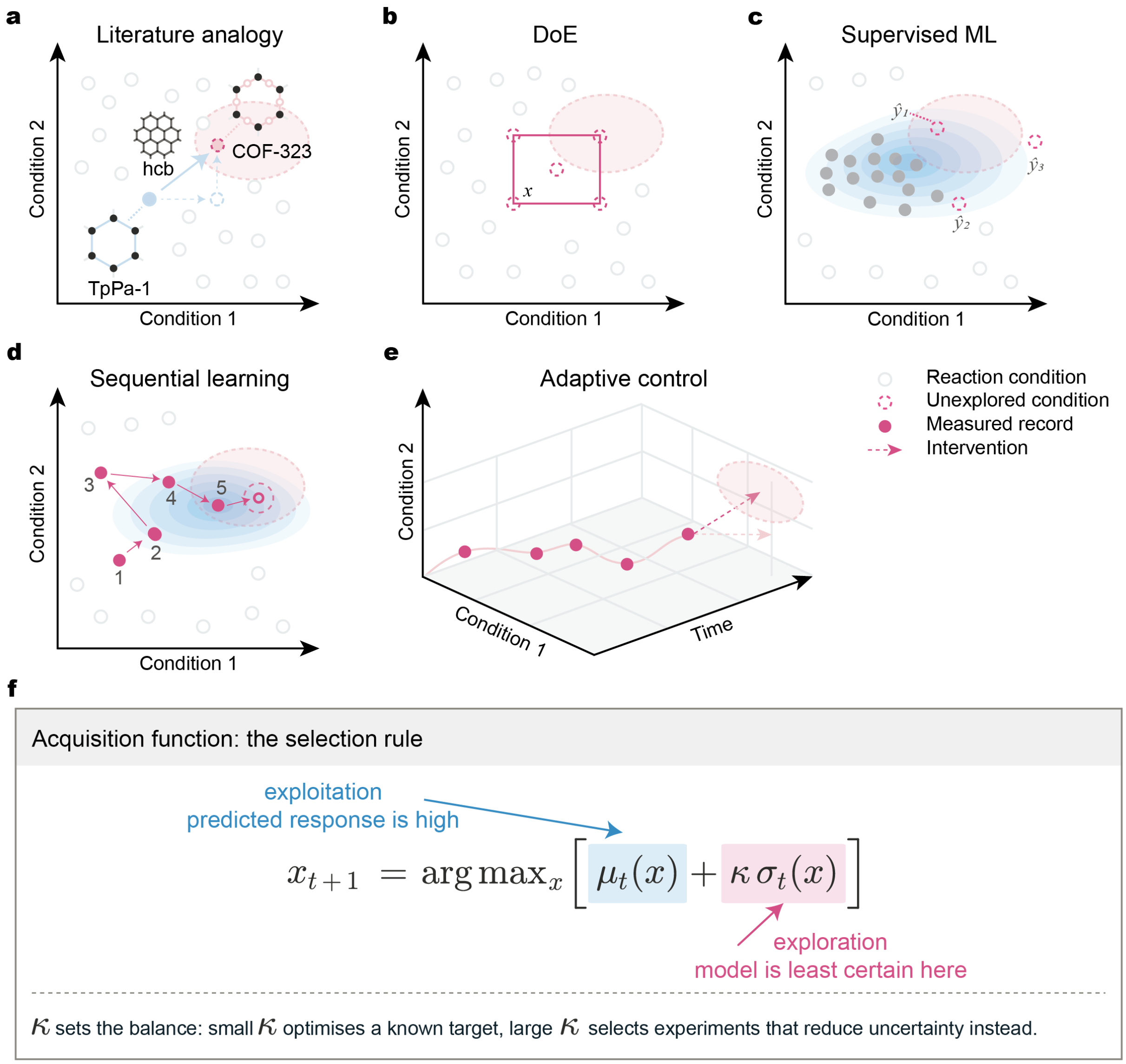

**Figure 3.** Prevailing strategies for exploring synthesis conditions. (a) Literature analogy with one-factor-at-a-time variation. (b) Design of experiments, in which the key decision lies in defining the initial factor ranges (the design box). (c) Supervised learning, which interpolates across a broader synthesis space on the basis of learned chemical intuition. (d) Sequential learning, which adopts an iterative strategy that introduces a temporal dimension and converts prediction into decision-making, such that its output is an experiment rather than a numerical value. (e) Adaptive learning, which operates within a single experiment and steers its overall course through monitoring and control at each step. (f) Experiment selection rules, typically based on an acquisition function that balances exploitation of current knowledge with exploration of unknown regions and returns the condition at which this balance is maximized.

The first algorithmic step is to convert local experimental judgment into a quantitative map, namely, to establish how variations in synthesis conditions relate to variations in framework outcome. In its simplest form, this can be achieved through design of experiments (DoE), in which a defined synthesis region is systematically sampled to determine the effects of individual variables and their interactions on an experimentally measured response.[158,159] This contrasts with the traditional practice of starting from a literature analogue for a one-factor-at-a-time progression, in which a single reaction parameter is varied while the others are held fixed (**Figure 3a**).[40–45] Factorial designs, full or fractional,[160,161] instead vary factors jointly (e.g., metal-to-linker ratio, modulator amount, solvent fraction, and temperature) to distinguish their individual effects from their interactions. Response-surface designs, such as central composite[162] and Box–Behnken arrangements,[163] can then resolve curvature and locate an interior optimum. Thus, the most critical reaction parameters leading to the intended phase with high crystallinity are narrowed to a few, which can then be mapped together across a local domain. Representative studies on framework materials such as Fe-MIL-88A [$Fe_3O(Fum)_3$; $Fum^{2-}$ = fumarate],[50] Y-UiO-66 [[$Y_6(\mu_3\text{-}OH)_8(BDC)_5(OAc)_{1.5}]^{1.5-}$; $OAc^-$ = acetate],[49] MIL-101 [$M_3O(BDC)_3$; M = Cr, V, or Mn],[51] COF-366 [$(TAPP)(BDA)_2$, imine-linked; TAPP = 5,10,15,20-tetrakis(4-aminophenyl)porphyrin, BDA = 1,4-benzenedicarboxaldehyde],[164] and SNW-1 [$(Mel)_2(Tpal)_3$, imine-linked; Mel = melamine, Tpal = terephthalaldehyde][165] collectively illustrate that DoE maps the region sampled experimentally to resolve condition effects and find a local, interpretable description; by itself, nevertheless, that map remains bounded by the factor levels chosen at the outset (**Figure 3b**).

Once that region has been sampled, the next algorithm-guided decision concerns the expected outcome of an unmeasured condition. In machine learning, supervised models learn $x \mapsto y$, where $x$ can encode reaction parameters (e.g., precursors, stoichiometry, solvent composition, modulation, temperature, time, operations), while $y$ can represent one or more experimentally observed outcomes (e.g., framework phase, crystallinity, permanent porosity, thermal stability) (**Figure 3c**). Considering that most MOF and COF campaigns presently comprise on the order of $10^2$–$10^3$ experimentally comparable records for a given predictive task,[46–48,166–171] classical machine-learning methods are well suited to mapping such relationships from the available data without the data demands of more parameter-rich models. For example, tree ensembles such as random forests[172] and XGBoost[173] are useful for the reason that they capture nonlinear relationships among mixed variables—categorical variables such as solvent or modulator identity and continuous variables such as concentration, temperature, and time—while feature analyses expose associations that remain chemically testable. An early example in HKUST-1 [$Cu_3(BTC)_2$; $BTC^{3-}$ = 1,3,5-benzenetricarboxylate] quantified the relative importance of synthesis variables as a measure of chemical intuition.[174] This concept has since been extended to larger experimental records of MOF and COF synthesis in which building-unit and reaction parameters are related to crystallization outcomes,[14,17,119,175] with success–failure records further defining both sides of the synthesis landscape.[21,121] Beyond fully supervised learning, in the event that failure records are scarce, one can consider positive-unlabeled learning, in which known successes are considered alongside conditions of unknown outcome.[120,176] In this way, synthesis conditions and framework outcomes can be represented as relationships, and unmeasured outcomes can be proposed through prediction.[177,178] This train-to-predict paradigm is prospective; nevertheless, it naturally raises a further question: which unmeasured condition should be tested next? Answering this question brings experiment selection itself into the learning of chemical intuition.

The third algorithmic consideration is sequential learning, whereby the outcome of each experiment is fed back to determine which condition should be attempted next (**Figure 3d**). By virtue of quantified uncertainty, experiment selection can balance exploration of poorly understood conditions with exploitation of those predicted to be favorable so as to reduce human bias (i.e., the tendency to repeatedly favor familiar regions of synthesis space).[157,179] In laboratory practice, this takes the form of an iterative cycle in which each measured outcome informs the selection of the next condition. To this end, Bayesian optimization[56–58] provides a means of formalizing this cycle through a surrogate model ("response proxy"), which serves as a probabilistic stand-in for the synthesis response, and an acquisition function ("selection rule"), which uses predicted performance and uncertainty to determine where to experiment next (**Figure 3f**). An early demonstration in ZIF synthesis used 36 iterations to improve the crystallinity of ZIF-67 [$Co(MeIM)_2$; $MeIM^-$ = 2-methylimidazolate].[122] In MOFs, Bayesian optimization was demonstrated in the discovery of previously unreported synthesis conditions for MOF-321 [Al(OH)(PZVDC); $PZVDC^{2-}$ = pyrazole-3,5-vinylenedicarboxylate] and MOF-322 [Al(OH)(TVDC); $TVDC^{2-}$ = thiophene-3,5-vinylenedicarboxylate] used rapid microwave synthesis to identify high-crystallinity conditions typically within 10–20 three-reaction iterations, in essence allowing chemical intuition for each new framework to be learned from scratch with accumulated failures.[23] The same work demonstrated its generality in COF-323, which in turn achieved 43% higher BET surface area than originally reported,[23,100] and comparable strategies have since been applied to other COF systems, for example, COF-284 [(TAB), phenyl-linked; TAB = 1,3,5-trifluoro-2,4,6-tris(4-acetylphenyl)benzene][166] and SUM-99 [(Tp)(Pa-HA), β-ketoenamine; Pa-HA = 2,5-diamino-1,4-benzenedihydroxamic acid].[123] As such, quantified uncertainty sequentially guides experiments between unexplored and favorable regions of MOF and COF synthesis space, through which chemical intuition can be progressively learned.

On the basis of the foregoing discussion, it is instructive, however, to separate reaction optimization from learning and discovery. In predictive synthesis, the former seeks conditions that improve a defined outcome for a known MOF or COF, whereas the latter seeks experiments that reveal previously unresolved relationships or expose new regions of synthesis space. This distinction is illustrated by the ZIF-A family,[21] where crystallization relationships learned from known imidazolate linkers were iteratively extended to previously unexplored linkers, with each experimental outcome refining which linker–condition combinations were most likely to yield new crystalline frameworks, such as ZIF-A6 [$Zn(BrIM)_2$; $BrIM^-$ = 5-bromoimidazolate] and ZIF-A9 [$Zn(mhbIM)_2$; $mhbIM^-$ = 6-methoxybenzimidazolate]. What such algorithmic discovery ultimately seeks to resolve can be appreciated from systematically mapped synthesis landscapes. Across rare-earth trimellitate MOFs, for example, variation of metal identity and reaction conditions delineated distinct regions for MIL-81 [RE(trimel)($H_2O$); RE = Sc, Y, or La–Lu; $trimel^{3-}$ = trimellitate], MIL-92 [$RE_2$(trimel)$(H_2O)_2$], and NU-2501 [RE(trimel)$(H_2O)_3$], together with regions yielding mixtures or no crystalline product.[168] A related strategy applied iterative learning across 520 possible building-unit combinations, leading to PL-COF-11 [(ETTA)(BTD-TBA), imine-linked; ETTA = 1,1,2,2-tetrakis(4-aminophenyl)ethene, BTD-TBA = 4,4′,4″,4‴-((benzo[c][1,2,5]thiadiazole-4,7-diylbis(4,1-phenylene))bis(azanetriyl))tetrabenzaldehyde] as a highly fluorescent COF.[180] In this manner, the fundamental objective progresses from identifying conditions favorable to crystallization toward establishing the boundaries that separate one framework outcome from another and how their domains change with chemistry.

Having such domains in hand raises the prospect of formalizing an important feature of chemical intuition, namely, the ability to transfer what is learned from one synthesis landscape to another. It should be highlighted that, for MOFs and COFs, systematic changes in node identity, building-unit geometry, linkage chemistry, and topology provide related, albeit chemically distinct, synthesis problems. Algorithmically, such knowledge can be transferred and organized through (i) multi-task models,[179,181,182] in which related synthesis problems are learned jointly so that experimental information acquired for one target informs predictions for another—for example, simultaneously learning crystallization outcomes for several linkers within a ZIF family; (ii) transfer learning,[183–185] in which relationships established from data-rich historical experiments are used to initialize and accelerate learning for a new, data-poor system, including related framework families (e.g., from Zr- to Hf-MOFs or imine- to hydrazone-linked COFs) and across synthetic approaches (e.g., solvothermal, microwave, sonochemical, or mechanochemical routes); and (iii) meta-learning,[186–188] in which experience accumulated across many related tasks is used to learn how a model should adapt when only the first few observations of a new system are available. The extent to which these approaches can identify relationships that persist across framework families will ultimately determine which empirical observations can be elevated to general principles of MOF and COF synthesis.

These developments, from structured experimentation and static prediction to sequential learning and ultimately transfer across chemical systems, together broaden the role of algorithms in synthesis. In sum, chemical intuition becomes measurable when a proposed experimental choice is accompanied by an expected outcome, calibrated uncertainty, and a chemical reason that can be tested. Each experiment then evaluates not only the prediction but also the synthesis relationship on which it is based.[30,63,137] As experimental evidence accumulates, the confidence, limits, and generality of such relationships can be progressively established, allowing chemical intuition to be expressed as experimentally testable knowledge in lieu of remaining implicit in individual experience. The subsequent step is to close the experimental loop such that these propositions are continually confronted with observation.

**Table 1.** Representative algorithm-guided approaches to MOF and COF synthesis.

| Representative framework(s) | Algorithm or model | Decision mode[a] | Input | Output | Data size[b] | Ref. |
|---|---|---|---|---|---|---|
| Y-UiO-66 | Fractional factorial; ANOVA[c] | DoE | Synthesis conditions | Yield | 16 conditions (duplicates) | 49 |
| Fe-MIL-88A | Central composite design | DoE | Synthesis conditions | Particle size | 20 experiments (8 factorial + 6 axial + 6 center) | 50 |
| MIL-101 (Cr, V, Mn) | Taguchi L9[c] orthogonal array; ANOVA | DoE | Synthesis conditions; metal type | Adsorption capacity; selectivity | 27 conditions (3 metals × L9; triplicates) | 51 |
| MIL-100(Fe)/COF (Mel–Tpal[c]) | Definitive screening design; extreme vertices mixture design; ANOVA | DoE | Precursor molar ratios | Nitrogen content; $CH_4$ uptake | 26 conditions (13 DSD[c] + 13 mixture; 7 tested for $CH_4$) | 52 |
| nanoCOF-366 | Full factorial design | DoE | Reagent concentrations | Size; morphology | 9 conditions (3 × 3) | 164 |
| Cu/COF (Mel–Tpal) | Full factorial design ($3^2$); extreme vertices mixture design; ANOVA | DoE | Precursor ratios; temperature; time; stirring speed | Amine loading; COF yield | 49 conditions (9 + 9 + 13 + 18) | 165 |
| MOF-520; ZIF-8; MOF-808 | RF[c] | Static prediction | Synthesis conditions | Crystal form | 764 records | 14 |
| $COF_{BPQD\text{-}TPDA}$; $COF_{TAPy\text{-}PMDA}$ | XGBoost[c] | Static prediction | Monomer fingerprints | Synthesis conditions | 587 records (2 validation COFs) | 119 |
| ZIF-8; ZIF-67; Co-MOF-74 | PU[c] neural networks | Static prediction | Synthesis conditions | Crystallinity score | 1,003,748 records (3,748 positive + $10^6$ unlabeled) | 120 |
| ML-COF-1 to -8 | RF | Static prediction | Monomer fingerprints | Synthesis conditions | 4,425 records (2,603 successes + 1,822 failures) | 121 |
| ZIF-A5; ZIF-A6; ZIF-A9 | Greedy selection (RF/GP[c]) | Sequential | Linkers; conditions | Framework formation | 4,416 experiments (2,688 training + 1,728 replay) | 21 |
| MOF-321; MOF-322; COF-323 | BO[c] (RF; EI[c]); GPT-4 | Sequential | Synthesis conditions | Crystallinity | 322 experiments (3 frameworks) | 23 |
| ZIF-67 | BO (RF; EI) | Sequential | Synthesis conditions | Crystallinity | 48 experiments (12 seed + 36 BO) | 122 |
| TAPB-PDA; SUM-99 | BO (GP; EI) | Sequential | Synthesis conditions | Crystallinity; phase selectivity | 164 experiments (129 seed + 35 BO) | 123 |
| ZIF-67 | BO; kinetic fitting | Sequential | Conditions; optical kinetics | Nucleation rate; extent of reaction | 33 vials 35 vials benchmark | 126 |
| COF-284 | BO (RF; EI) | Sequential | Synthesis conditions | Crystallinity | 160 experiments (80 seed + 80 BO) | 166 |
| Cu-HKUST-1; Zn-HKUST-1 | GA[c]; RF analysis | Sequential | Synthesis conditions | Crystallinity; porosity | 90 experiments (3 GA generations) | 174 |
| PL-COF-10; PL-COF-11 | DFT[c]-pretrained MLP[c]; Siamese network | Sequential | Electronic descriptors | Photoluminescence quantum yield | 22 COFs (520 candidates) | 180 |
| POMOF1-1; POMOF1+3 | XGBoost; uncertainty sampling | Sequential | Synthesis conditions | Crystallization boundaries | 107 records (42 seed + 65 assigned negatives) | 189 |
| Al-PMOF | GA | Sequential | Synthesis conditions | Crystallinity; yield | 45 experiments (25 seed + 20 GA) | 190 |

[a] Decision modes classify how the experiments or predictions in each study were selected.

[b] Data size refers to the corpus or campaign stated in the cited source.

[c] Abbreviations: ANOVA, analysis of variance; BO, Bayesian optimization; DFT, density functional theory; DoE, design of experiments; DSD, definitive screening design; EI, expected improvement; GA, genetic algorithm; GP, Gaussian process; L9, nine-run orthogonal array; Mel, melamine; MLP, multilayer perceptron; PU, positive-unlabeled; RF, random forest; Tpal, terephthalaldehyde; XGBoost, extreme gradient boosting.

## Closing the Experimental Loop

The process of predictive synthesis begins with a chemically grounded expectation of what a chosen set of building units might form and ends only when the resulting observation has altered the course of the experiment that follows. Realizing the broader prospect of programming MOFs and COFs on demand to address societal needs therefore requires closing this experimental loop,[30,32] such that prediction, synthesis, and feedback are joined into an autonomous discovery cycle that learns from each experiment (**Figure 4a**).[191–193] Such a cycle connects (i) prior chemical knowledge that establishes what might be attempted, as derived from literature precedent; (ii) algorithmic decision-making that determines what should be attempted, as in Bayesian optimization; (iii) experimental execution that realizes the proposed synthesis, as through robotic platforms; and (iv) characterization that establishes what has formed and redirects the next action, as through PXRD feedback, rendering the cycle self-correcting. The significance of connecting these elements lies in the progressive accumulation of experimental observations, through which relationships between synthesis conditions and framework outcomes can be quantified and transferred, ultimately providing a means of formalizing chemical intuition.

The starting point of such a cycle is the chemical knowledge accumulated before an experiment is carried out. One form can be considered a range-type prior (i.e., "*under what conditions have related frameworks formed*"), where literature analogues delimit plausible solvent, modulator or catalyst, temperature, and stoichiometric ranges for a new synthesis.[14,194,195] A second can be defined as a property-related prior (i.e., "*how have related frameworks been observed to behave*"), where accumulated observations associate chemical features with properties and functions.[196–198] A third can accordingly be established as a configuration-constrained prior (i.e., "*what framework can these building units construct*"), where the geometry and connectivity of the chosen building units constrain the topologies and configurations considered plausible for their assembly.[199–202] On this basis, such information constitutes the synthesis prior of MOFs and COFs, in other words, the chemical expectation available before an experiment is performed and ultimately tested against observation.

Once such a synthesis prior has been established, the challenge is to translate chemical expectation into an action that can be experimentally carried out in matter. One can imagine an LLM-based agent serving as the "computational brain" of a physical laboratory, where chemical knowledge is converted into experimental action.[20,151] It is worthwhile remarking that, although the LLM and LLM-based agent or agentic system are sometimes used interchangeably,[203] by definition, an agent fundamentally acts autonomously on a user's behalf toward a specified goal, making it particularly suited to the autonomous discovery cycle described herein. In this role, in lieu of directly predicting synthesis outcomes, an agent can query literature and databases of MOFs and COFs and, subsequently, invoke algorithmic or computational tools for prediction, uncertainty quantification, generation, and simulation,[32] before converting the resulting decision into instructions for physical action by laboratory instruments. Building on earlier multi-agent automated experimentation for MOF-321, MOF-322, and COF-323,[23] agentic systems have further incorporated generative and quantum-mechanical models, which led to the experimental validation of five MOFs, including AI-MOF-4 [$Zn_4(DMFum)_3 \cdot 2(DMF)$; $DMFum^{2-}$ = 2,3-dimethylfumarate, DMF = *N*,*N*-dimethylformamide].[204] More recently, this cycle has been augmented with automated characterization and quantitative feedback, enabling the *de novo* discovery of COF-2000 [(Tp)(TABTA), β-ketoenamine; TABTA = $N^1$,$N^3$,$N^5$-tris(4-aminophenyl)benzene-1,3,5-tricarboxamide] through high-throughput robotic synthesis and PXRD.[195]

In addition, closing the experimental loop requires that a decision generated digitally can be rendered faithfully in matter.[205] Automation provides the physical counterpart to algorithmic decision-making by encoding experimental actions in standardized form, such that addition order, timing, thermal history, sample identity, and post-synthetic treatment can be reproduced and each observation remains attached to the experiment that produced it.[189,190] One can further imagine two or more geographically separated "cloud laboratories" exchanging the same digital protocol,[206,207] where standardized liquid and solid handling allows the cross-laboratory MOF or COF synthesis to be reconstructed with reduced human-dependent variation and difference in procedure completeness. Notwithstanding, most MOF and COF laboratories are brownfield environments, where insoluble products must pass through existing reactors, washing and activation operations, and characterization instruments.[2,3,144,155] In such settings, two complementary philosophies can be envisioned. One is a machine-centered approach (i.e., "adapt experiment to machine"), where synthesis and characterization are reformulated around dedicated robotic workcells, standardized vessels, and instrument interfaces amenable to automated handling (**Figure 4b**).[23,122,189,195] The other is a human-centered approach (i.e., "adapt machine to experiment"), where humanoid robots equipped with vision and dexterous manipulation operate existing laboratory infrastructure—gloveboxes, fume hoods, analytical balances, and vacuum ovens whose dimensions, access points, and controls were originally designed around human operation.[208,209] Both routes ultimately seek reproducible execution coupled to traceable observations, providing the physical embodiment of predictive synthesis algorithms and agents in the laboratory.

Beyond automated execution, closing the experimental loop also requires a characterization strategy capable of providing the appropriate observation at the appropriate experimental time and resource (**Figure 4c**). Indeed, in many MOF and COF workflows, synthesis can already be parallelized more readily than structural characterization,[155,205] such that the analytical bottleneck progressively replaces the action of preparing more reaction in quantities. By analogy to multifidelity computational modeling,[27] predictive synthesis can be naturally viewed as a multifidelity experimental problem, where inexpensive proxy measurements are acquired broadly and progressively more informative measurements are reserved for cases where they alter a chemically consequential decision.[210] A camera, for example, can follow dissolution, cloudiness, precipitation, crystal appearance, and morphology across hundreds of reactions,[168,211,212] which can identify clear solutions, trace product, and single crystals to prioritize samples for diffraction, as illustrated by computer-vision workflows and robotic microscopy of Co-MOF-74 [$Co_2$(DOT)].[211]

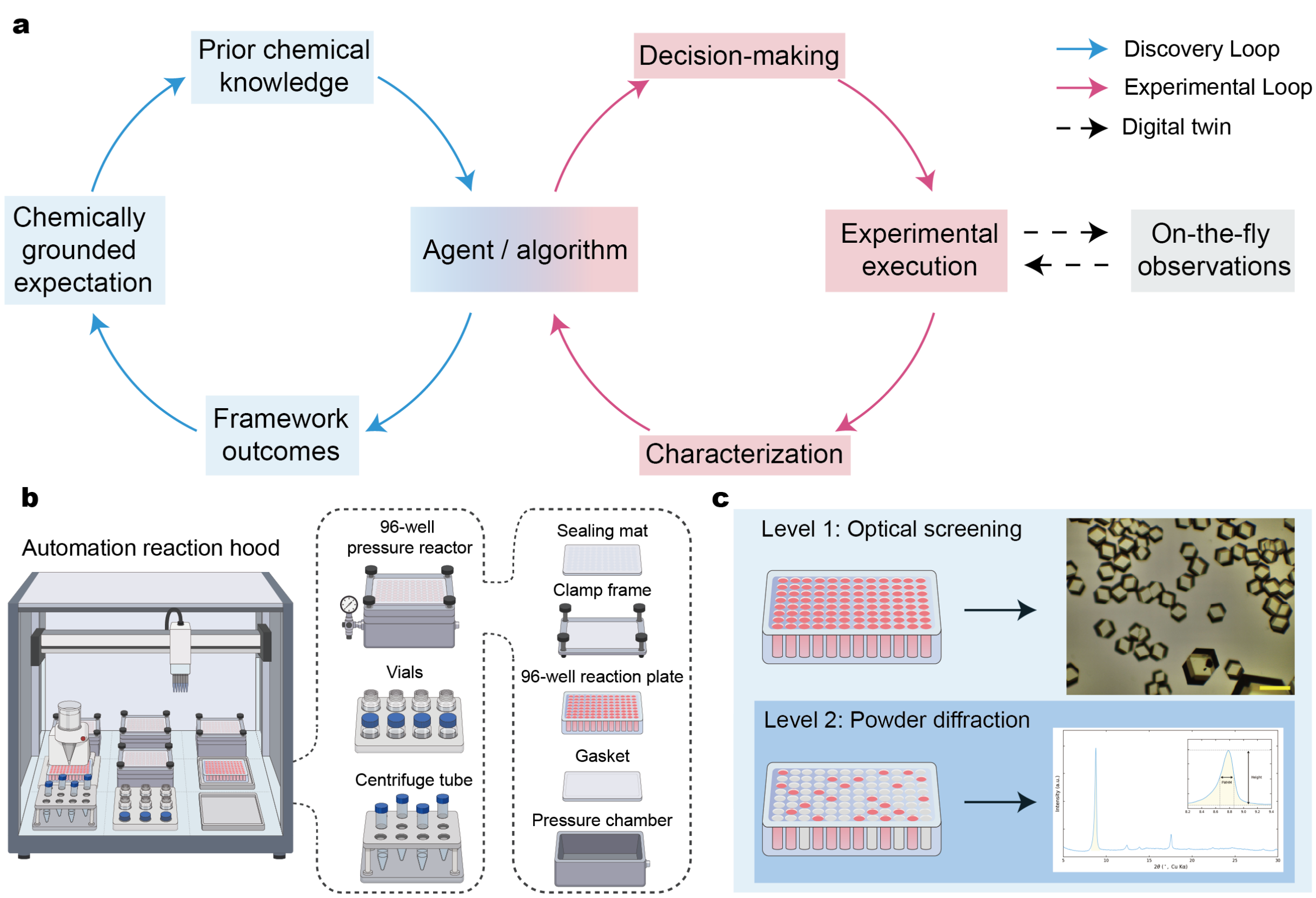

**Figure 4.** Schematic of the closed experimental loop. (a) The autonomous discovery cycle begins with a chemically grounded expectation and proceeds through iterative experimentation directed by an agent or algorithm, yielding experimental outcomes that inform subsequent decisions. (b) High-throughput automated reaction setup adaptable to diverse experimental settings. (c) Hierarchical characterization strategy in which analytical fidelity is progressively increased over the course of the experiment.

At the second level, PXRD can distinguish known or poorly crystalline products, leaving unmatched or structurally distinct materials for single-crystal XRD or electron diffraction. Such a hierarchy is already emerging in ZIF and rare-earth MOF synthesis,[21,168] where optical observations precede progressively more resolved diffraction measurements. Conceptually, the same principle can also be extended to COFs, where inline optical or vibrational measurements could guide subsequent PXRD and selected structural or property characterization. Hence, multimodal and multifidelity feedback does not require every experiment to receive every measurement; the value lies in progressively matching analytical fidelity to the structural and chemical uncertainty revealed by the experiment.

At its core, predictive synthesis depends on how rapidly an observation can be reconciled with the preceding action and effectively translated into a chemically justified intervention. Hitherto, autonomous synthesis cycles have generally taken an episodic form, with each experiment carried to an endpoint and characterized before the next is selected.[191,192] At the current stage, while a static loop can sufficiently learn where a framework forms, in the long run it cannot intervene while precursor conversion, nucleation, growth, and phase transformation are still taking place, processes whose time dependence is intrinsic to MOF and COF crystallization.[95,106–108] A complementary form of the closed experimental loop would preserve the traditional autonomous discovery cycle between experiments while augmenting it with on-the-fly observations acquired during the reaction and continuously reconciled with the predictive model through a reaction-level "digital twin,"[126,213,214] such that feedback arrives in time to alter the course of the same, ongoing synthesis. For instance, an unexpectedly early onset of precipitation could trigger reduced microwave or sonication power or an earlier quench, whereas delayed formation could lead to an altered thermal profile, extended reaction time, or reagent addition. Precedents from related crystallization systems have shed light on how time-resolved feedback can enable analogous experimental control.[124,125] The ability to learn from classical endpoint feedback, coupled with adaptive temporal feedback during the reaction, allows the experimental loop to operate across both synthesis space and synthesis time. In this manner, predictive synthesis advances from identifying the conditions under which a framework forms to directing the course by which it forms, bringing framework formation closer to design when synthesis can be proposed prospectively, observed as it unfolds, and redirected before its outcome is settled.

**Remaining Unknowns and Emerging Opportunities**

MOF and COF chemistry has reached a stage where the limiting unknown is the extent to which a learned relationship remains valid beyond the conditions from which it arose. In particular, there remains a disparity between what predictive models reproduce within a familiar synthesis space and what chemists require across unseen framework families and experimental settings. We describe relationships that retain predictive validity under such changes as “synthesis invariants.” Their establishment requires prospective tests on (i) unseen building-unit combinations; (ii) unseen metals, linkages, or topologies; (iii) external laboratory environments; (iv) changes in scale and equipment; and (v) incomplete observations. Interlaboratory variation and the scarcity of independent repeat syntheses show that a condition set may appear reliable within one record while remaining fragile as transferable knowledge.[40,129] Literature precedent should likewise not be treated as uniformly reliable,[40,129] and one opportunity is to classify synthesis records by experimental trustworthiness before they are used as prior knowledge for prediction. Against this background, community benchmarks should evaluate prospective target formation, uncertainty calibration, out-of-domain recognition, and cross-laboratory transfer using provenance-linked synthesis trajectories.[32,207,215–218] To this end, standard reporting formats and machine-actionable records provide the foundation for such comparisons.[2,135,143] Accordingly, a governing principle should transcend a particular framework or experimental setting, such that predictive synthesis progresses from local prediction toward universally transferable relationships between synthesis and framework formation.

An important extension of this generalization challenge appears when predictive synthesis progresses from laboratory discovery toward large-scale synthesis and commercialization. A synthesis carried out in a milliliter-scale vial cannot be assumed to remain the same synthesis after multiplication of reagent quantities, since mixing, heat transfer, reactor geometry, and downstream handling progressively become part of the synthesis trajectory.[34,219–222] In spite of successful demonstrations of MOF production at substantially increased scale, for example, CALF-20 [$Zn_2(Trz)_2(Ox)$; $Trz^-$ = 1,2,4-triazolate, $Ox^{2-}$ = oxalate],[37] scale-up remains strongly dependent on the chemistry and process by which a framework is formed.[223,224] The challenge for predictive synthesis, in turn, is to quantitively map which synthesis trajectories are intrinsically amenable to scale-up and which experimental variables must be transformed as scale increases. It is envisioned that, as the available protocols continue to grow, small- and large-scale protocols for the same framework could provide paired evidence for learning such relationships, thereby bringing process scale into the synthesis space itself.

Equally unresolved is the chemical meaning of failure in MOF and COF crystallization. A model trained only to identify the intended framework may assign comparable value to no precipitate, an amorphous product, and a competing crystalline phase, although these outcomes carry fundamentally different information. Success–failure learning has already shown the importance of preserving negative experimental evidence,[23,121,136,137,225] while phase mapping and rigorous structure determination reveal that an unintended crystalline product can contain a precise account of another path available to the reaction.[96,138,168] A discrepancy between prediction and resolved product should therefore be treated as a test of the synthesis relationship itself. The missing factor may arise from precursor chemistry, nucleation, or a later stage of activation. A model could then specify which intervention should redirect the trajectory and what evidence would invalidate that claim. Accordingly, assigning chemical meaning to unintended products is an integral part of predictive synthesis and a necessary condition for learning the boundaries separating one framework outcome from another.

An additional unknown—perhaps less obvious, but more fundamental—concerns how framework topology itself is communicated to a predictive model. In small-molecule chemistry, a SMILES string encodes atom identities and bonding according to a defined molecular graph;[226,227] by contrast, the three-letter topology symbols widely used in MOF and COF chemistry are assigned identifiers for periodic nets.[9,77,78] Some retain semantically interpretable, as in **sod** for sodalite, **hcb** for honeycomb, and **kgd** for kagome dual; others, including symbols such as **qom** and **ucb**,[228,229] are considerably less informative from the characters themselves, since topology symbols can originate from historical associations, institutions, or abbreviations introduced when the net is named. Consequently, representing a framework simply by its building units and a topology token can obscure the associated structural meaning, and a predictive model may naively memorize that certain building units frequently accompany a particular token without learning the underlying periodic graph.[20] This limitation becomes more fundamental considering that, up to now, topology diversity has remained relatively underexplored[230] compared with the much greater attention given to building-unit variation enabled by isoreticular principles,[81] with the power of topology as the synthesis (in)variant consequently underestimated. Remarkably, it represents a challenge unique to crystalline framework materials, for which a topology-aware representation should connect building units to the periodic graph they construct, in lieu of treating topology as a categorical label whose meaning must be inferred from precedent.

At the longest horizon, predictive synthesis of MOFs and COFs should expand from one-shot prediction of crystallization conditions toward multi-step inverse synthesis, where a targeted framework is traced backward to its most basic building units, and then forward again through an experimentally realizable synthesis trajectory. Structural inverse design already progresses from desired properties toward candidate frameworks, while topology analysis and synthesizability models constrain which proposed constructions are chemically plausible.[24,25,28,199] The difficulty is inherently many-to-many, as a single topology may be realized by several chemically distinct combinations of nodes and linkers, while the same combination of building units may give rise to more than one topology as precursor state or assembly kinetics change.[8,201] Once the framework has been deconstructed to its most basic molecular components, the problem begins to meet the domain of small-molecule retrosynthesis, where target molecules are recursively transformed into simpler precursors through sequences of chemically feasible steps.[231–234] For crystalline framework structures, however, these molecular routes must ultimately be joined again with the collective processes of nucleation and crystallization that select the extended structure. A useful planner should have the capability of connecting framework topology to building units, building units to accessible molecular precursors, and those precursors to the sequence of experimental operations by which the intended phase might emerge. Each step constrains what follows, while observations acquired during synthesis can redirect the proposed path when the chemistry departs from expectation.[154,235–237] Besides, reaction-level digital twins could add strength to this workflow by reconciling these trajectories with time-resolved observations as crystallization proceeds.[108,113,126,213] At that point, the shift is from a building-unit-centered view, in which molecular components and their properties are used to construct extended solids, toward a route-centered view, in which a targeted solid is connected backward to accessible chemicals and forward through the sequence of operations required for its formation. Such retrosynthesis planning also creates an opportunity to incorporate solvent use, precursor availability, energy demand, and waste generation into the synthesis objective,[238] bringing predictive synthesis of MOFs and COFs closer to a practice in which material realization and sustainability are considered together.

As predictive capability improves, equal attention must be given to reflecting on what constitutes a prediction itself within the scheme of synthesis. A prospective synthesis prediction, by its own nature, acquires scientific meaning only when it precedes the experimental outcome against which it is evaluated.[239–241] If a new MOF or COF is first synthesized by rational design and the model, prompt, candidate space, or training information is subsequently adjusted until that outcome becomes favored, the resulting exercise becomes post hoc rationalization even when its final form appears predictive. A related ambiguity may arise when an ostensibly unseen target has a nearly identical precedent within the information available to the model, yet that relationship remains unrecognized. Such considerations are especially relevant to generative AI,[242] where repeated prompting and model selection can progressively expose the researcher to the desired answer before a final prediction is declared, while benchmark contamination and data leakage can further inflate apparent model capability.[32,151,243–245] A credible predictive record should preserve and faithfully report the target, available knowledge, model state, and confidence before experiment, so as to distinguish prospective inference from retrospective reconstruction. Prospective and time-aware evaluation are useful in establishing the extent to which a model reaches into new chemistry in lieu of reproducing familiar precedent.[241,246,247] The broader opportunity is to move from individual predictions toward governing principles, where transfer tests generality, unintended phases reveal missing chemistry, inverse planning connects targets to realizable routes, and digital twins examine these relationships as synthesis unfolds.

## Concluding Remarks

Up to now, the chemistry of MOFs and COFs has progressed from a scarcity of isolated synthesis examples to an abundance of chemical space, whose composition and pore environment can be designed with a degree of precision uncommon among extended solids.[95] Notably, what follows from this is thus to bring equivalent precision to their formation. In this Perspective, we have outlined how synthesis evidence, algorithmic decision-making, and experimental feedback can be taken as a whole to render prospectively testable hypotheses and quantify chemical intuition. As a matter of fact, the importance of this approach lies in connecting individual synthesis attempts into a body of chemical evidence, through which the conditions leading to formation, failure, and transfer can be distinguished.[31,205] In this way, empirical synthesis is envisioned to be progressively transformed from a sequence of isolated outcomes into a means of establishing experimentally testable principles of framework formation.

Predictive synthesis, as an emerging chemical discipline,[62,63,238] will be realized when the conditions governing framework formation can be (i) proposed prospectively; (ii) tested quantitatively; and (iii) altered in response to the evolving experiment. At this juncture in crystalline porous materials, the question is what new chemistry such control will render possible. The answer may lie in multivariate MOFs and COFs,[248] whose ordered backbones can support many arrangements of chemical functionalities, some of which may act together to produce properties unattainable from the individual components.[249] Predictively directing their incorporation and spatial arrangement would transform this vast synthetic complexity into a programmable form, ultimately allowing sequence-dependent functions to be encoded within crystalline frameworks.

More broadly, predictive synthesis offers a means to broaden global access to the practice of MOF and COF chemistry. By placing experimentally established synthesis–structure–property relationships within accessible data-driven workflows, chemists worldwide could pursue targeted crystalline structures without independently reconstructing the same empirical knowledge.[30] In the fullness of time, these algorithms, when adopted as toolkits in routine laboratory practice, hold the potential to reduce bias and error in experimental decisions and open the vast synthesis space of MOFs and COFs to a broader community of chemists, in turn expanding our collective capacity to discover new framework chemistry with predictive confidence.

AUTHOR INFORMATION

**Corresponding Author**

**Zhiling Zheng** − *Department of Chemistry; Institute of Materials Science and Engineering, Washington University, St. Louis, Missouri 63130, United States;* orcid.org/0000-0001-6090-2258

**Authors**

**Bin Feng** − *Department of Chemistry, Washington University, St. Louis, Missouri 63130, United States;* orcid.org/0009-0003-1642-9776

**Reid A. Coyle** − *Department of Chemistry, Washington University, St. Louis, Missouri 63130, United States;* orcid.org/0009-0007-8888-729X

**Shyam Chand Pal** − *Department of Chemistry; Institute of Materials Science and Engineering, Washington University, St. Louis, Missouri 63130, United States;* orcid.org/0000-0003-3743-2571

**Peter Walther** − *Department of Chemistry, Washington University, St. Louis, Missouri 63130, United States*

**Saeun Park** − *Department of Chemistry, Washington University, St. Louis, Missouri 63130, United States*

**Author Contributions**

The manuscript was written through contributions of all authors. All authors have given approval to the final version of the manuscript.

**Conflicts of interest**

There are no conflicts to declare.

ACKNOWLEDGMENT

Z.Z. acknowledges funding and extends thanks for the support provided by the EQT Foundation through the Breakthrough Science Grant (No. 059994), Research Corporation for Science Advancement (RCSA) through the Scialog Collaborative Awards (SA-AUT-2026-073C and SA-AUT-2026-074A), Thinking Machines through the Tinker Research Grant, the Royal Society of Chemistry through the Researcher Collaborations grant (GR0042769), and Washington University in St. Louis through the Startup Grant. B.F. is grateful for the support from Washington University in St. Louis through the Fossett Fellowship.

TABLE OF CONTENTS

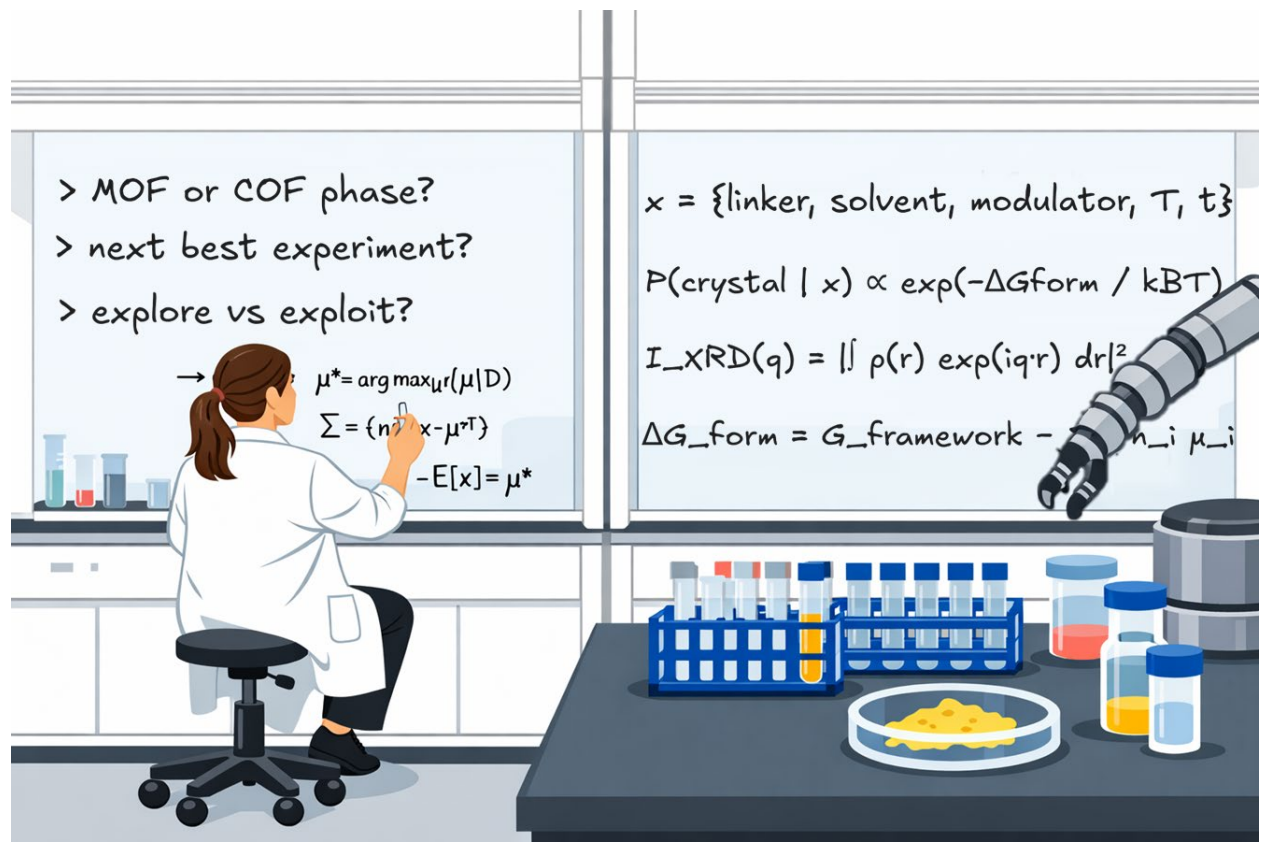